\documentclass[aps, twocolumn,superscriptaddress, 
amsmath,amssymb,reprint,numbers,noeprint
]{revtex4-1}

\usepackage[utf8]{inputenc}
\usepackage[english]{babel}

\usepackage{float}
\usepackage{graphicx}% Include figure files
\usepackage{dcolumn}% Align table columns on decimal point
\usepackage{bm}% bold math
\usepackage{amsmath}
\usepackage{color}
\usepackage{comment}
\usepackage{array}
\usepackage{siunitx}
\usepackage{comment}
\usepackage{makecell}
\usepackage{threeparttable}
\usepackage{textgreek}

\usepackage{multirow}

\usepackage{hyperref}
\hypersetup{
     colorlinks   = true,
     linkcolor    = blue,
     citecolor    = blue,
     urlcolor = blue
}

\newcolumntype{P}[1]{>{\centering\arraybackslash}p{#1}}

\definecolor{col1}{rgb}{0,0,0}
\newcommand{\sri}[1]{{\color{col1} #1}}
\definecolor{col2}{rgb}{0, 0, 0}

\definecolor{col3}{rgb}{0,0,0}

\definecolor{col4}{rgb}{0, 0, 0}

\begin{document}

\title{A window into NV center kinetics via repeated annealing and\\ spatial tracking of thousands of individual NV centers}

\author{Srivatsa Chakravarthi}
\email{srivatsa@uw.edu}
\affiliation{Department of Electrical and Computer Engineering, University of Washington, Seattle, Washington 98195, USA}
\author{Chris Moore}
\affiliation{Department of Physics, University of Washington, Seattle, Washington 98195, USA}
\author{April Opsvig}
\affiliation{Department of Electrical and Computer Engineering, University of Washington, Seattle, Washington 98195, USA}
\author{Christian Pederson}
\affiliation{Department of Physics, University of Washington, Seattle, Washington 98195, USA}
\author{Emma Hunt}
\author{Andrew Ivanov}
\author{Ian Christen}
\affiliation{Department of Physics, University of Washington, Seattle, Washington 98195, USA}
\author{Scott Dunham}
\affiliation{Department of Electrical and Computer Engineering, University of Washington, Seattle, Washington 98195, USA}
\author{Kai-Mei C Fu}
\affiliation{Department of Electrical and Computer Engineering, University of Washington, Seattle, Washington 98195, USA}
\affiliation{Department of Physics, University of Washington, Seattle, Washington 98195, USA}

\date{\today}% It is always \today, today,

\begin{abstract}
Knowledge of the nitrogen-vacancy (NV) center formation kinetics in diamond is critical to engineering sensors and quantum information devices based on this defect. Here we utilize the longitudinal tracking of single NV centers to elucidate NV defect kinetics during high-temperature annealing from 800-1100\;$^\circ$C in high-purity chemical-vapor-deposition diamond. We observe three phenomena which can coexist: NV formation, NV quenching, and NV orientation changes. Of relevance to NV-based applications, a 6 to 24-fold enhancement in the NV density, in the absence of sample irradiation, is observed by annealing at 980\;$^\circ$C, and NV orientation changes are observed at 1050\;$^\circ$C. With respect to the fundamental understanding of defect kinetics in ultra-pure diamond, our results indicate a significant vacancy source can be activated for NV creation between 950-980\;$^\circ$C and suggest that native hydrogen from NVH$_y$ complexes plays a dominant role in NV quenching, \sri{supported by} recent {\it ab initio} calculations. Finally, the direct observation of orientation changes allows us to estimate an NV diffusion barrier of \sri{4.7 $\pm$ 0.9~eV}.  
\end{abstract}
\maketitle

\section{Introduction}
Nitrogen-vacancy centers in diamond are point defects that are utilized for sensing~\cite{ref:rondin2014mnv, ref:degen2017qs, ref:schirhagl2014nvc} and quantum information applications~\cite{ref:childress2013dnv, humphreys_deterministic_2018} due to their long spin coherence time~\cite{balasubramanian_ultralong_2009} and optically accessible spin states~\cite{doherty_nitrogen-vacancy_2013}. Of significant interest are methods to synthesize negatively charged nitrogen-vacancy (NV$^-$) centers with minimal perturbation to the local environment, preserving the quantum and optical properties of the center. Full control over NV$^-$ center formation requires a detailed understanding of all the underlying defect kinetics. While there has been progress in understanding NV$^-$ center energetics~\cite{mainwood_nitrogen_1994,goss_vacancy-impurity_2005,deak_formation_2014,antonov_statistical_2014} and engineering NV$^-$ center formation kinetics~\cite{favaro_de_oliveira_tailoring_2017}, here we show how a new tool, the tracking of thousands of individual NV$^-$ centers, lends insight into the formation, quenching, and orientation kinetics of a quantum defect in an ultrapure host.

We perform repeated vacuum annealing of chemical vapor deposition (CVD) diamond, and track individual NV$^-$ centers between anneals via photoluminescence confocal microscopy on a large experimental volume ($350\times350\times25$~\textmu m$^3$). We show a 6-fold to 24-fold sample- and location-dependent increase in the overall concentration of NV$^-$ centers due to annealing at 980\;$^\circ$C. This  indicates there is a source of vacancies within CVD diamond at moderate temperatures that enables one to significantly increase NV$^-$ density without introducing additional lattice damage via irradiation. Individual NV$^-$ tracking allows us to go beyond measuring simple net increases and decreases. Coincident with appearances, we are able to observe significant NV$^-$ disappearances, with peaks in disappearances near 960\;$^\circ$C and 1050\;$^\circ$C. The disappearances near 960\;$^\circ$C would be completely masked by appearances in ensemble studies\sri{,} but their presence provides an essential clue regarding the quenching process at higher temperatures.

At 1050\;$^\circ$C, we observe large-scale NV$^-$ center orientation changes. After the first 1050\;$^\circ$C anneal, orientation changes are observed in the absence of a significant number of disappearances or appearances. Thus, only partial dissociation occurs at these temperatures. This lack of full dissociation indicates the observed disappearances at 960\;$^\circ$C and 1050\;$^\circ$C are not due to NV-disassociation but due to a quenching source (e.g. H). Finally, beginning at temperatures above 1000\;$^\circ$C, we begin to observe almost complete NV quenching originating from the surface. 

\section{Experiment}

\subsection{Samples}
Our studies are performed on \sri{five} commercial CVD samples (Element Six). Samples A, B, \sri{D and E} are ``electronic grade'' and have a substitutional nitrogen concentration of [N$_s$] $<1$ ppb, according to the manufacturer. Sample C has [N$_s$] $<1$ ppm. All samples have a \{100\} crystal orientation. Details of the growth conditions of the samples are not provided by Element Six, however the growth temperature of similar samples is around 830\;$^\circ$C~\cite{ref:isberg2002hcm}. Prior studies of Element Six CVD diamond samples indicate a ratio of 1000~:~3 for [N$_s$]~:~[NV$^-$] for \{100\}-oriented growth surfaces~\cite{edmonds_production_2012}. Using this ratio and confocal NV$^-$ imaging, we estimate [N$_s$]$_{\textrm{A,B,D,E}}\approx$1~ppt in \sri{electronic grade} samples and [N$_s$]$_\textrm{C}\approx$50~ppb in sample C (Table~\ref{tab:Samples}).

\begin{table*}[]
  %  \begin{minipage}{10 cm}
    \begin{threeparttable}[t]
\centering
\def\arraystretch{2}
\begin{tabular}{|P{1.1cm}|P{1.8cm}|P{1.8cm}|P{1.8cm}|P{2cm}|P{1.4cm}|P{1.6cm}|P{1.4cm}|P{1.4cm}|}
\hline
\multirow{2}{*}{Sample} & Initial [N$_s$] estimate & Initial [NV$^-$]$_i$ & Maximum [NV$^-$]$_{max}$ & Enhancement $\displaystyle \frac{\mathrm{[NV^-]_{max}}}{\mathrm{[NV^-]_{i}}}$ & Surface depletion & Orientation changes & Large scans & Depth scans \\ \hline
A\tnote{1} & 0.8 ppt &  2.3 ppq &  2.7 ppq &  1.2\tnote{2} & Yes & Yes & Yes & No \\ \hline
B & 1.2 ppt &  3.7 ppq & 24.4 ppq &  6.6 & Yes & Yes & Yes & Yes \\ \hline
C & 50 ppb  & 166 ppt  & 4 ppb &  24.1 & No & en\tnote{3} & No & Yes \\ \hline
D\tnote{4} &  1.5 ppt  & 4.5 ppq  & 16.4 ppq & 3.7 & Yes & NA & No & Yes \\ \hline
E\tnote{4} &  0.6 ppt  & 1.9 ppq  & 13.2 ppq & 6.9 & Yes & NA & No & Yes \\ \hline
\end{tabular}
\begin{tablenotes}
\item[1] Sample underwent 2-hour anneals at 800\;$^\circ$C, 900\;$^\circ$C, 1000\;$^\circ$C and 1100\;$^\circ$C.
\item[2]  Enhancement not observed due to short annealing times and inhomogeneous surface driven depletion~\cite{ref:supp_mat}.
\item[3] Ensemble measurement, orientation changes are not distinguishable.
\item[4] Samples only annealed one time at 970\;$^\circ$C for 150 h. 
\end{tablenotes}
\end{threeparttable}
%\end{minipage}
\caption{\sri{Summary of all samples with corresponding substitutional nitrogen, initial NV$^-$ and maximum NV$^-$ densities. NV$^-$ densities in samples A, B are estimated from the large-area scan data sets. For sample C-E densities are estimated from the depth scan data sets. Sample A,B,D,E densities are calculated from counting single NV$^-$ centers in the confocal images and dividing by the scan volume (area $\times$ depth of focus). Details on the conversion is provided in the Supplemental Materials~\cite{ref:supp_mat}. Sample C density is estimated by normalizing the PL intensity to that of a single NV$^-$ center and dividing by the confocal volume.}}
\label{tab:Samples}
\end{table*}

\subsection{Annealing}
The diamond samples were annealed under vacuum ($<$~1e-7 mbar) with anneals conducted in order of increasing temperature. A ramp time of 2 hours was used. To minimize surface fluorescence, the samples were cleaned for 90 minutes in a fuming \sri{acid bath (initial composition 1:1:1  H$_2$SO$_4$:HNO$_3$:HCLO$_3$)} maintained at 250\;$^\circ$C prior to annealing. 

Sample A was annealed for 2 hours at temperatures 800\;$^\circ$C, 900\;$^\circ$C, 1000\;$^\circ$C and 1100\;$^\circ$C and probed at 96~\textmu m below the surface. In this data set, a large number of NV$^-$ appearances (25\;\% of the total population) and disappearances (20\;\%) were observed at 1000\;$^\circ$C. At 1100\;$^\circ$C, disappearances dominated with the total NV$^-$ density depleted by more than 80\;\%~\cite{ref:supp_mat}.

The sample A study motivated a finer temperature study from 950-1050\;$^\circ$C performed further from the surface. This full study was performed on sample B with annealing temperatures and times found in Table~\ref{tab:Anneals}. \sri{Due to the timescale of the observed processes, it was not feasible to reach thermal equilibrium at each temperature. Instead, the decision to increase the annealing time or annealing temperature was qualitatively based on the magnitude and/or saturation behavior of the observed changes. } Annealing of sample C began at 980\;$^\circ$C in order to reproduce the NV$^-$ appearances observed in sample B. \sri{Samples D and E were annealed only once at \sri{970}\;$^\circ$C.}

\subsection{NV identification and tracking}
\begin{figure}
\centering
\includegraphics[width=3.4in]{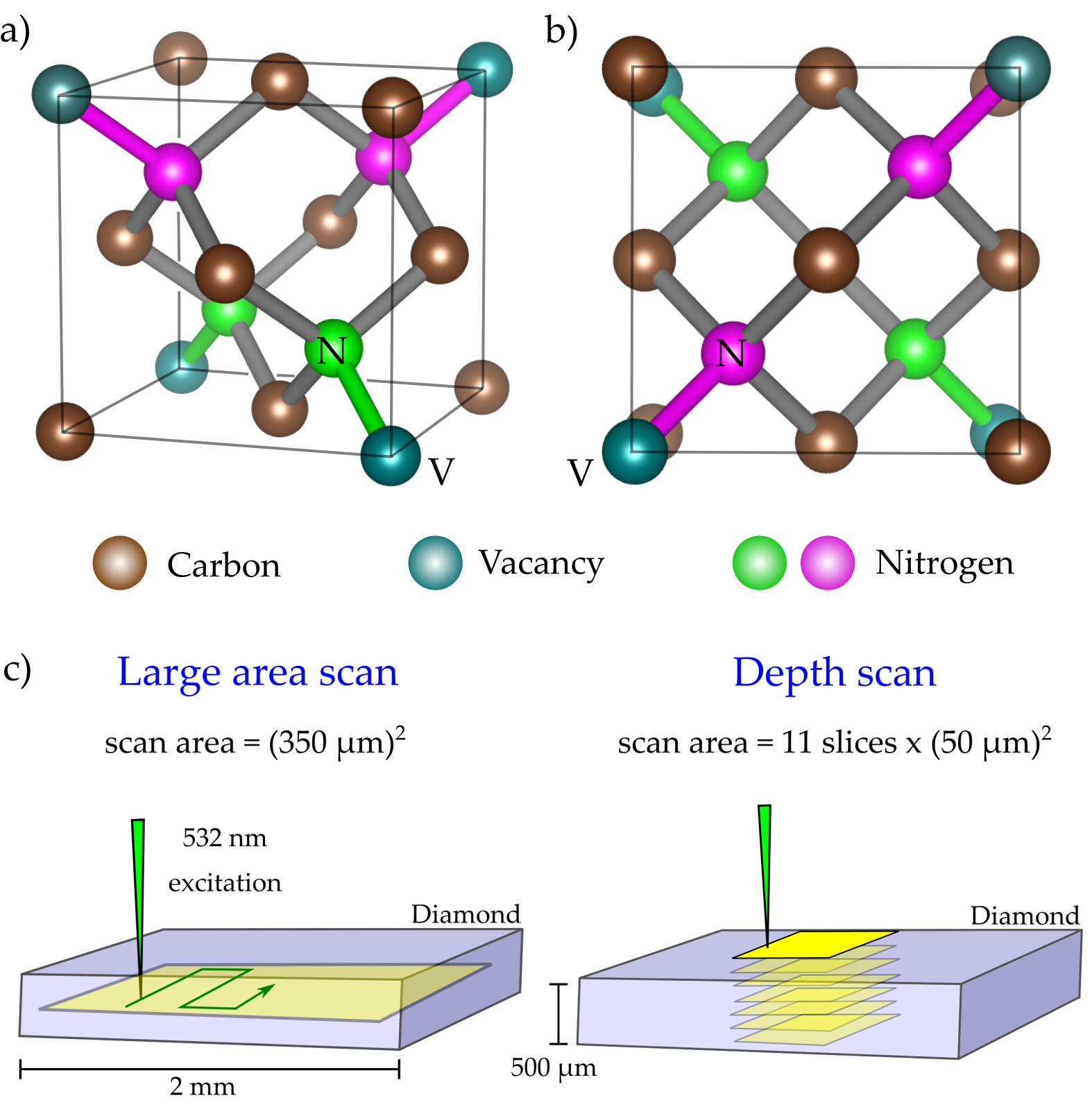}
\caption{The two optically distinguishable NV$^-$ orientation sets are encoded in green and magenta. (a) side view. (b) top view. All surfaces are  \{100\} planes. \sri{(c) Illustration of large-area scan and depth scan geometries.}}
\label{fig:nv_orientations}
\end{figure}

\begin{figure*}[t]
\centering
\includegraphics[width=6.3in]{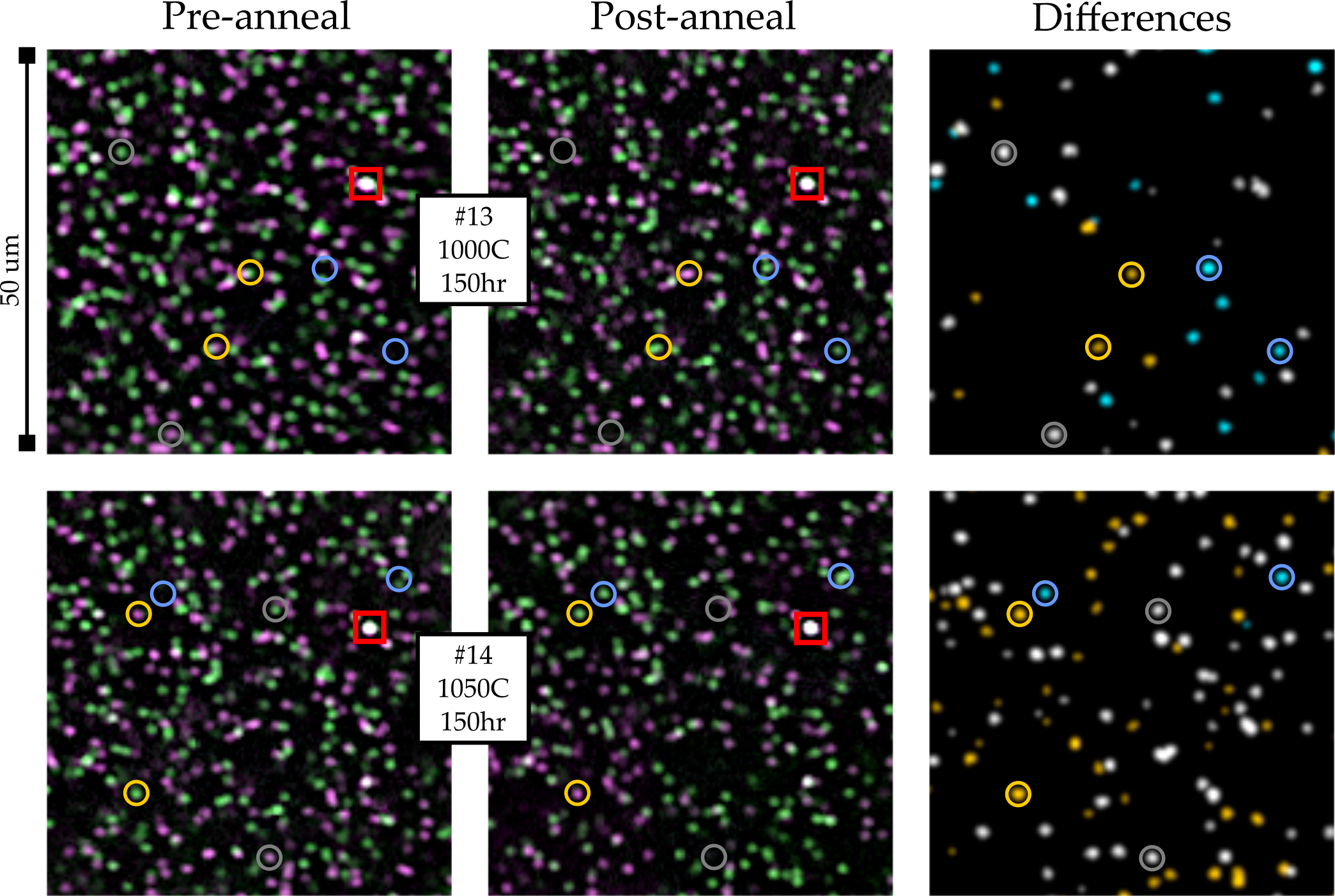}
\setlength{\belowcaptionskip}{-10pt}
\caption{A sequence of subareas of the ``large-area scan'' confocal images showing NV$^-$ changes. \sri{Green and magenta indicate the two sets of distinguishable NV$^-$ orientations.} Two examples each of appearances, disappearances and re-orientations are indicated by blue, grey and yellow circles, respectively. \sri{The red square marks a persistent defect used for image alignment.}}
\label{fig:comparison}
\end{figure*}

NV$^-$ centers were imaged via confocal microscopy \sri{utilizing a 0.75 NA objective}. The centers were non-resonantly excited with a linearly-polarized 532~nm laser. The NV$^-$ phonon sideband emission was filtered (660-800$~$nm) and detected with an avalanche photodiode. \sri{Further experimental details are provided in the Supplemental Material SM~\cite{ref:supp_mat}}. The excitation polarization angle was used to probe NV$^-$ center orientation~\cite{edmonds_production_2012,ref:alegre2007pse}. We can distinguish two sets of orientations, with two orientations in each set, in our \{100\}-oriented diamond samples. The two sets are indicated as green and magenta in Figs.~\ref{fig:nv_orientations} and \ref{fig:comparison}. An automated stage with tandem micrometers and precision piezo-actuators enabled the generation of precise spatial maps of \sri{each} NV$^-$ center in the experimental area. 

To characterize the outcome of each anneal, we performed two types of measurements: ``large\sri{-}area scans'' and ``depth scans'' \sri{, as illustrated in Fig.~\ref{fig:nv_orientations}c}. Large\sri{-}area scans longitudinally tracked individual NV$^-$ centers in the same ($350\times350\;$\textmu m$^2$) area at a depth of \sri{$240\;$}\textmu m ($96\;$\textmu m) from the top surface for sample B (A) with a depth-of-focus of 25~\textmu m. \sri{The top surface is the surface closest to the microscope objective (SM~\cite{ref:supp_mat})}. Large area scans captured both sets of NV$^-$ orientations. Image registration, aided by persistent luminescent defects and local NV$^-$ constellations, was performed to match individual NV$^-$ center locations. The encoded NV$^-$ orientation data provided additional information to confirm location and was utilized to identify orientation changes. The NV$^-$ center appearances, disappearances and orientation changes were obtained through image processing and validated manually. Image processing details are provided in the Supplemental Material~\cite{ref:supp_mat}. Examples of NV$^-$ appearances, disappearances, and orientations in confocal images can be seen in Fig.~\ref{fig:comparison}.  

Depth scans allow us to sample the NV$^-$ density through the vertical cross-section of our diamond samples. A confocal scan $(50\times50\;$\textmu m$^2$) was performed in five different regions in a quincunx pattern with $500\;$\textmu m spacing. A single excitation polarization is used and depth scans do not track individual NV$^-$ centers. For sample C, in which single NV$^-$ centers could not be detected due to the high NV$^-$ density, only depth scans were performed. Depth scans capture the NV$^-$ density variation across and through the sample.

\section{Results}
\begin{figure*}
\centering
\includegraphics[width=7in]{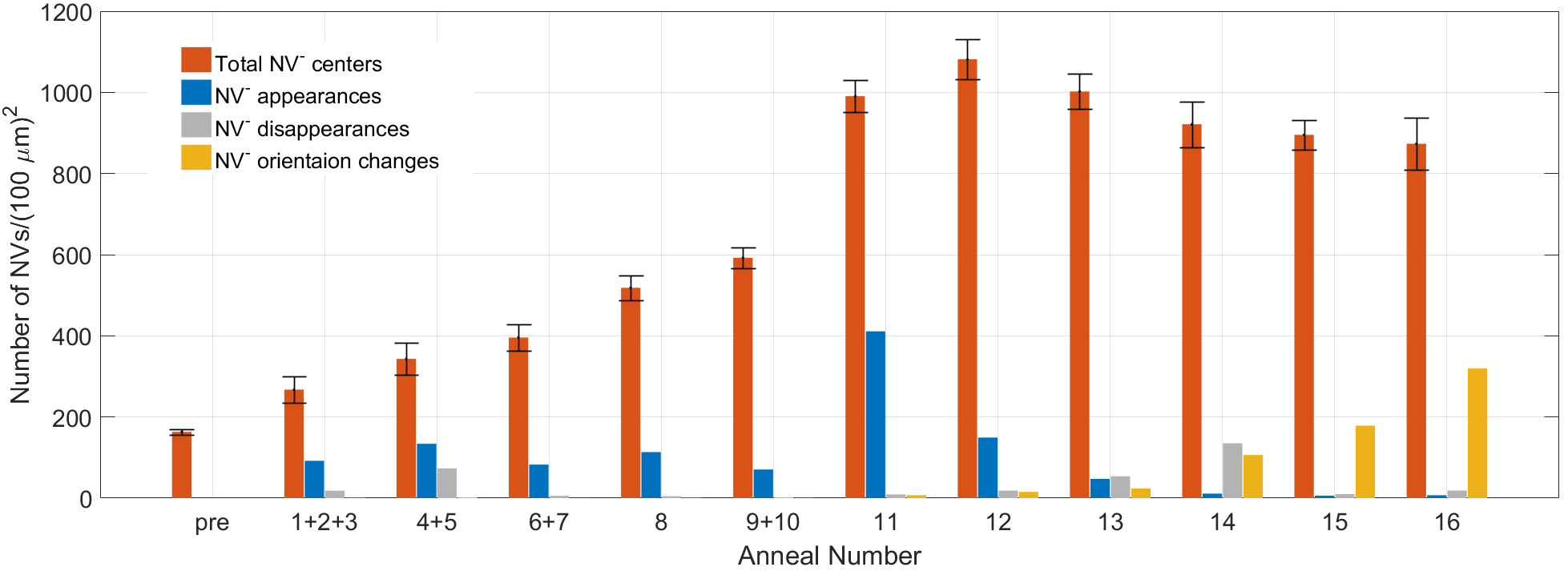}
\setlength{\belowcaptionskip}{-10pt}
\caption{Total NV$^-$ density and observed changes obtained from the large-area scans ($\approx$ 350 $\times$ 350 \textmu m$^2$) of sample B after every anneal. \sri{Several anneals are combined due to an inability to accurately match a significant portion of the scan area in the intermediate scans. For combined anneals, comparisons are made before and after the first and last anneal. Error bars represent uncertainty due to automation errors and differences in scan area between datasets~\cite{ref:supp_mat}}. Note the significant increase in annealing time which occurs at anneal 11.}
\label{fig:largescan}
\end{figure*}

The results from both the large area scans (sample A) and depth scans (samples A,B) are summarized in Figs.~\ref{fig:largescan} and \ref{fig:depth_ppm_cvd19}, \sri{respectively} and in Table~\ref{tab:Samples}. 

\vspace{1mm}
{\it Appearances}:
In sample B, we observe the NV$^-$ center density increase during each anneal going from 950\;$^\circ$C to 980\;$^\circ$C \sri{(anneals 3-10 in Table~\ref{tab:Anneals} and Fig.~\ref{fig:largescan})}. The continual observation of NV$^-$ appearances motivated longer anneals (150 hours) starting at 980\;$^\circ$C \sri{(anneal 11)} to observe saturation effects. A lower NV$^-$ creation rate is indeed observed in the second 980\;$^\circ$C 150 hour anneal \sri{(anneal 12)}. A maximum 6-fold increase is observed after this anneal in the large area scan data \sri{(pre-anneal v. anneal 12, Fig.~\ref{fig:largescan} and Fig.~S2~\cite{ref:supp_mat}}), while the averaged data from the depth scans show a maximum 8-fold increase \sri{(pre-anneal v. anneal 15, Fig.~\ref{fig:depth_ppm_cvd19}a)} . The variation in both density and density changes are attributed to the spatial variation of defect densities incorporated during the growth process. \sri{Average NV$^-$ density varied by a factor of three across the five depth scan regions.}

\vspace{1mm}
{\it Disappearances}: \sri{There are two different types of disappearance events: a spatially homogeneous NV$^-$ depletion and a surface-driven inhomogeneous depletion.} In sample B at 250~\textmu m below the surface, we observe a small number of disappearances, homogeneously throughout the scan region, initially at every anneal temperature increment. We observe a larger number of disappearances at 960\;$^\circ$C and 1050\;$^\circ$C (Fig.~\ref{fig:largescan}). Very few disappearances are observed after the first 1050\;$^\circ$C anneal, suggesting the source of the quenching has been depleted. In the sample C ensemble measurements, we can also visibly observe a decrease in NV$^-$ density during the first 1050\;$^\circ$C anneal (Fig.~\ref{fig:depth_ppm_cvd19}c). 

A different behavior was observed at 1100\;$^\circ$C in sample A. At this temperature, we observed the majority of NV$^-$ centers at a depth of 40~\textmu m disappear. This surface-driven inhomogeneous depletion is also observed in sample B, as shown in Fig.~\ref{fig:depth_ppm_cvd19}a; similar disappearances toward the surface begin to happen around $T\sim980$\;$^\circ$C (anneal 10) ~\sri{and extend more that 100 microns into the surface by anneal 16.} 

\vspace{1mm}
{\it Orientation changes}: Significant orientation changes are observed at 1050\;$^\circ$C. As shown in the sample B data (Fig.~\ref{fig:largescan}), the orientation changes are accompanied with an increase of disappearances during the first 1050\;$^\circ$C anneal. However, in subsequent 1050\;$^\circ$C anneals, orientation changes are not accompanied by a significant number of disappearances or appearances. Curiously, we observe a significant increase in orientation changes during \sri{subsequent} 1050\;$^\circ$C anneal\sri{s}.

\section{Discussion}

{\it Appearances}:
\sri{We discuss three candidate processes for NV$^-$ center appearances: charge state conversion of NV$^0 \rightarrow$ NV$^-$, direct dissociation of NVH $\rightarrow$ NV + H, and diffusion and capture of single vacancies, N+V $\rightarrow$ NV. 

In high-purity type IIa diamond, both the neutral and negatively charged state of NV centers are stable under optical excitation.  The ratio of the two populations at a given Fermi-level depends on the excitation wavelength and intensity~\cite{ref:siyushev2013ocs,ref:aslam2013pii}. In agreement with other groups studying high-purity CVD diamond with 532~nm excitation~\cite{ref:alsid2019pda, ref:chen2013omc}, we find NV$^-$ is the dominant charge state ~\cite{ref:supp_mat} in pre- and post-annealed samples. 

If NV is formed by the dissociation of a larger complex, the homogeneity and scale of NV$^-$ appearances indicate this dissociating species should be uniform and abundant. In similar CVD samples, the NVH density is 10 times greater than the NV$^-$ density~\cite{edmonds_production_2012}, suggesting NVH as the NV source. However, the dissociation of NVH $\rightarrow$ (NV + H) can be ruled out as this complex is observed to be stable until 1400\;$^\circ$C~\cite{ref:cruddace2007mro, khan_colour-causing_2013}. 

Single vacancies combining with native nitrogen is a third possible source. Single vacancies can exist in the neutral and negative (V$^0$ or V$^-$) charge states in diamond in which NV$^-$ is the dominant NV charge state. It is generally accepted that vacancies are only mobile in their neutral charge (V$^0$) state at temperatures above 700\;$^\circ$C, with an activation energy of 2.3~eV~\cite{davies_vacancy-related_1992}. However, it has been confirmed that the NV center forms as a unit during CVD growth~\cite{edmonds_production_2012, ref:osterkamp2019epa} and not via vacancy capture. Thus during growth, all vacancies incorporated are either in the V$^-$ state or trapped in complexes (V$_2$, NVH, VH, etc.). Divacancy (V$_2$) dissociation can be ruled out because it has been observed to anneal out at 800\;$^\circ$C~\cite{ref:twitchen1999epr} and other complexes have high ($>$1100\;$^\circ$C) dissociation temperatures. This leaves the negative vacancy (V$^-$). {\it Ab initio} calculations estimate a 3.4~eV migration energy for V$^-$~\cite{ref:breuer1995aii}. Assuming a 30 THz attempt frequency, this migration energy corresponds to a diffusion length on the order of 100~nm for a 150 hour anneal at 980\;$^\circ$C. While there is uncertainty in both the migration energy and attempt frequency, direct migration and trapping of V$^-$ appears to be a viable candidate for NV center formation.

We note that NV formation in the absence of irradiation has previously been reported between 1500-1600\;$^\circ$C~\cite{ref:luhmann2018sec,osterkamp_engineering_2019} in CVD diamond. The vacancy source in these reports could be similar to what we observe in our lower temperature, longer duration anneals, or could be due to dissociation of NVH~\cite{ref:cruddace2007mro, khan_colour-causing_2013} or VH~\cite{glover_hydrogen_2004}.}

\begin{table*}[t]
  \begin{tabular} {|| m{2.5cm} | m{0.75cm}m{0.75cm}m{0.75cm}m{0.75cm}m{0.75cm}m{0.75cm}m{0.75cm}m{0.75cm}m{0.75cm}m{0.75cm}m{0.8cm}m{0.8cm}m{0.8cm}m{0.8cm}m{0.8cm}m{0.8cm}  ||}
  \hline
    Anneal \# & 1 & 2 & 3& 4 & 5 & 6 & 7 & 8 & 9 & 10 & 11 & 12 & 13 & 14 & 15 & 16 \\
    \hline\hline
    Temperature,\;$^\circ$C & 800 & 800 & 950 & 960  & 960 & 970 & 970 & 970 & 980 & 980 & 980 & 980 & 1000 & 1050 & 1050 & 1050 \\
    Time, h   & 2 & 10 & 10 & 10 & 10 & 2 & 10 & 20 & 2 & 10 & 150 & 150 & 150 & 150 & 150 & 150 \\
    Sample & B & B & B & B & B & B & B & B & B & B & B,C & B,C & B,C & B,C & B,C & B,C \\
    \hline
  \end{tabular}
      \caption{Anneals conducted on samples B \& C.}
    \label{tab:Anneals}
\end{table*}

\begin{figure*}
\centering
\includegraphics[width=7in]{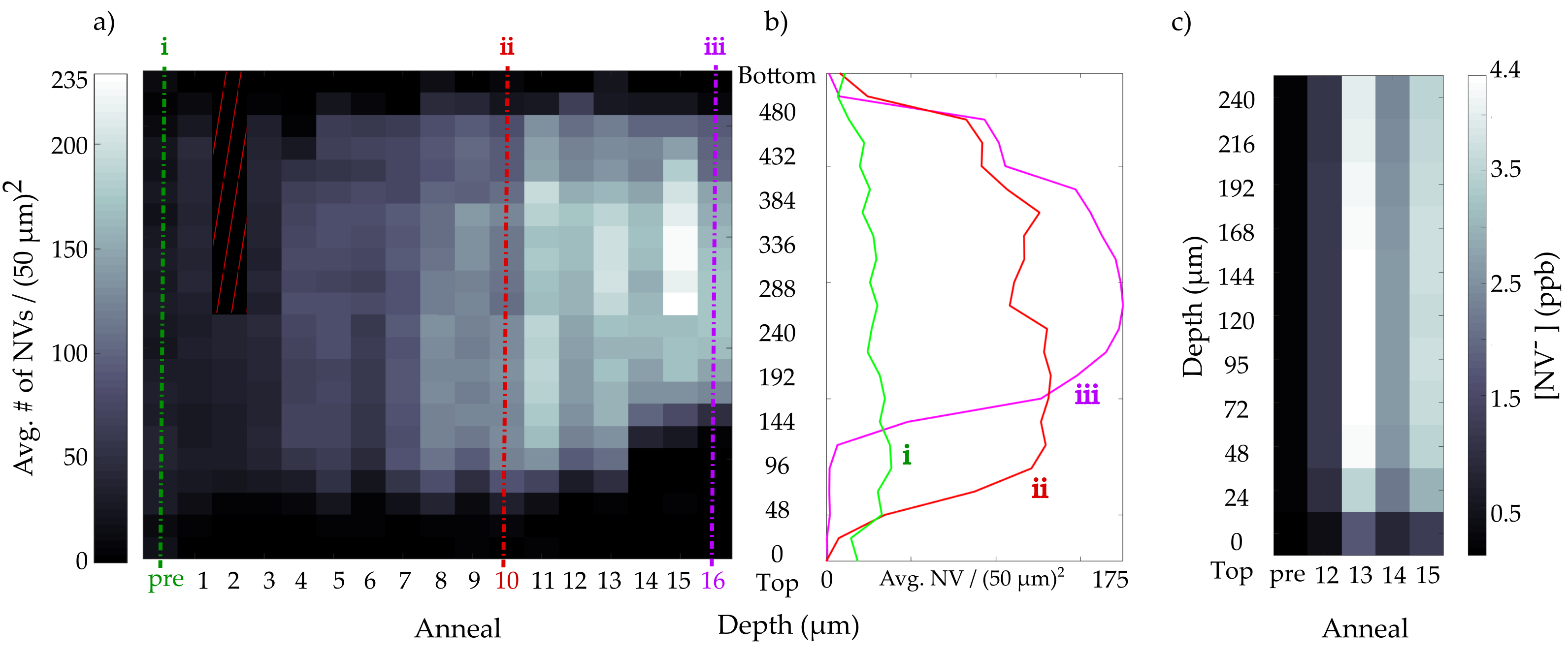}
\setlength{\belowcaptionskip}{-10pt}
\caption{(a) Plot of the average number of NV$^-$ centers per (50~\textmu m)$^2$ area in sample B as a function of sample depth and anneal. \sri{In order to access the full depth profile with the limited piezo scan range, the sample is imaged from both the top and bottom surfaces. The two data sets are stitched at $\sim\;$250~\textmu m}. \sri{The red hashed area indicates missing data.} (b) Line plots of slices notated in (a). (c) Plot of NV$^-$ density in sample C as a function of sample depth and anneal.}
\label{fig:depth_ppm_cvd19}
\end{figure*}

\vspace{1mm}
{\it Disappearances}:
\sri{The spatially homogeneous} disappearances occur at temperatures below the full NV$^-$ dissociation temperature (which our data show must be above 1050\;$^\circ$C), strongly suggesting another impurity is being trapped at the NV site. \sri{Hydrogen is the most likely candidate.} Given the abundance of C, N, H, and vacancies in our samples, complexes either in the form of CH \cite{fuchs_hydrogen_1995, goss_theory_2002}, NH \cite{goss_theory_2002}, V$_x$H$_y$~\cite{mehandru_hydrogen_1992,salustro_characterization_2018,khan_colour-causing_2013} or N$_x$VH$_y$~\cite{glover_hydrogen_2004, salustro_hydrogen_2018,khan_colour-causing_2013} are potential \sri{spatially homogeneous hydrogen sources}. 

A comparison of our experimental results and recent {\it ab initio} calculations indicate the NVH$_y$ complex, in which $y = 1,2,3$, is a plausible H source for the observed NV$^-$ quenching near 960\;$^\circ$C and 1050\;$^\circ$C. Salustro \textit{et al.}\sri{~\cite{salustro_characterization_2018}} calculated the dissociation energies, $E_H$, for the following reaction path,
$$
\mathrm{NVH}_3 \xrightarrow[\mathrm{R_1}]{\mathrm{E_H^1}} \mathrm{NVH}_2 + \mathrm{H} \xrightarrow[\mathrm{R_2}]{\mathrm{E_H^2}} \mathrm{NVH} + 2\mathrm{H} \xrightarrow[\mathrm{R_3}]{\mathrm{E_H^3}} \mathrm{NV} + 3\mathrm{H}
$$
finding $\mathrm{E_H^1} = 2.83$~eV, $\mathrm{E_H^2} = 3.17$~eV, and $\mathrm{E_H^3}= 3.68$~eV. The dissociation rate for a given reaction is given by $R_\mathrm{i}(T) = \nu_i\exp(-(\mathrm{E_H^i+E_D})/k_BT)$, \sri{in which $T$ is the annealing temperature (in K), $E_D$ is the dissociation barrier, and $\nu_i$ is the rate for capture of the  H by the NVH complex. The $\nu_i$ primarily depend on the hopping rate for intersitial H and can be expected to be nearly the same for all the complexes.  

If we assume $R_1$ (the rate at which NVH$_3\rightarrow$ NVH$_2$ + H) becomes \sri{discernible} at 960\;$^\circ$C (the temperature at which we first observe disappearances), we can estimate when $R_2$ become \sri{discernible} using $R_2(T_2) \approx R_1 (960\;^\circ$C), finding $T_2\approx 1110\;^\circ$C.  While this  is higher than our observed second peak in disappearances at 1050\;$^\circ$C, we note that the 1050\;$^\circ$C anneal was more than 7 times longer than the 960\;$^\circ$C anneal. Extrapolating further to the dissociation of NVH, we obtain that $R_3(1330\;^\circ$C) $\approx R_1 (960\;^\circ$C). $T_3 = 1330\;^\circ$C} is consistent with the experimental observation of NVH disappearance beginning near 1400\;$^\circ$C during shorter, 4 hour anneals~\cite{ref:cruddace2007mro,khan_colour-causing_2013}. 

We cannot use these annealing times to quantitatively predict when the ratios between $R_1, R_2$, and $R_3$ will become equal as disappearances typically saturate during each anneal. This saturation suggests a depletion of the H source. Qualitatively, however, the larger magnitude of disappearances at 1050\;$^\circ$C is consistent with a larger expected density [NVH$_2$] relative to [NVH$_3$].

In sample B, near-total NV depletion near the surfaces was observed starting at 980\;$^\circ$C (Fig.~\ref{fig:depth_ppm_cvd19}b), with the depletion layer growing in subsequent anneals. The surface dependence suggests a second, \sri{inhomogeneous, surface-driven hydrogen source. We believe this source is also the cause of the 80\;\% depletion in sample A at 1100~$^\circ$C~\cite{ref:supp_mat}, \sri{strongly suggesting large hydrogen diffusion length in diamond}. The observation of a surface-driven depletion suggests it may be desirable to protect the diamond surface to preserve near-surface NV centers during annealing.}

Both NV$^-$ PL quenching as well as enhancement has been reported in the literature over a wide range of temperatures from 1100-1500$^\circ$C~\cite{ref:tetienne2018spd, ref:yamamoto2013esc, ref:ozawa2018tsp,pezzagna_creation_2010}, with higher temperature quenching often attributed to NV migration to form larger complexes~\cite{pinto_diffusion_2012,pezzagna_creation_2010}.  Due to the short \sri{N$_s$} \sri{diffusion} length, complex formation \sri{with multiple N atoms (e.g. N$_x$VH$_y$, N$_x$V)} can only explain quenching in high-\sri{N$_s$} samples~\cite{ref:luhmann2018sec}.  Our results indicate \sri{the following factors could contribute to the observed variance in NV quenching reported in the literature}: (1) Compositional sample variance with respect to hydrogen traps, (2) Simultaneous creation and quenching events with the specific annealing temperatures and times determining a net outcome in either direction and (3) the hydrogen content in the annealing furnace, suggested by our observed surface dependence of the depletion layer. 

\vspace{1mm}
{\it Orientation changes}:
Recently, a loss of preferential orientation was observed in ensembles of oriented NV centers~\cite{ref:ozawa2018tsp} also at 1050\;$^\circ$C. The mechanism for this orientation change was believed to involve the full dissociation of the NV center with subsequent recombination with either the original N or another N in the lattice. Our single defect tracking shows that full dissociation does not \sri{occur} at 1050\;$^\circ$C. If the vacancy is able to migrate, we would expect to observe orientation changes accompanied by NV$^-$ disappearances.\sri{ Additionally,} if the vacancy diffusion length is comparable to the N$_s$-N$_s$ spacing ($\sim$1~\textmu m in sample B), we would also expect to see new NV$^-$ appearances. However, after an initial round of disappearances in the first 1050\;$^\circ$C anneal, we primarily only observe orientation changes. This suggests there is an attractive force, extending to at least the 3rd nearest neighbor (the minimum separation required for NV$^-$ re-orientation), between the vacancy and the nitrogen~\sri{\cite{pinto_diffusion_2012}}. While practical implementation may be challenging, this result opens a path toward preferential orientation by annealing under strain~\cite{karin_alignment_2014}.

\sri{If many \sri{re-orientation cycles} occur during a single anneal, we would statistically expect 50\;\% of the NV$^-$ centers will undergo a detectable orientation change. The maximum fraction of orientation changes we observe is \sri{37\;\%} during anneal 16\sri{, suggesting that} NV$^-$ centers do not undergo multiple \sri{re-orientation cycles} during each anneal. Thus, we can utilize the fractional rate from anneals 11-16 to estimate the orientation-change barrier $E_b$ using  the re-orientation rate $R = \nu \mathrm{exp}(-E_b/kT)$. A fit to the data gives $E_b =$ 4.7 $\pm$ 0.9~eV~\cite{ref:supp_mat}. This value \sri{is comparable} to the theoretical value of 4.85~eV~\cite{pinto_diffusion_2012} determined by density functional theory calculations. With a larger dataset and annealing times in which thermal equilibrium is reached for all processes, this method of observing orientation changes could be used to provide benchmarks for theoretical estimates of $\nu$ and $E_b$.} 

\vspace{1mm}
\sri{{\it Comparison between samples B and C}: Despite the almost five orders of magnitude difference in initial NV$^-$ density between samples B and C, the averaged results are qualitatively consistent, suggesting a similar interplay of defect species exists over a wide range of N doping. In both samples a large increase in the NV$^-$ density is observed with long anneals at 980\;$^\circ$C and a decrease in NV$^-$ density is observed at 1050\;$^\circ$C. While the results are qualitatively similar, quantitatively the magnitudes of the observed changes differ, suggesting the importance of the microscopic environment on the competing processes for defect formation. As we previously noted, variation in enhancement is observed even within a single sample. While beyond the scope of this work, future work could correlate annealing behavior with prior (NV$^-$\sri{, N$_s$ \& NVH$^-$}) density, appearances, or disappearances in this dataset~\cite{ref:DVN/V2EYN7_2019} or future datasets.} 

\section{Conclusions}
The formation and dissociation of observable defects has long been used to probe defect kinetics in crystals. For example, a model for vacancy diffusion in diamond was facilitated by correlating V$^0$ (GR1) emission, V$^-$ (ND1) absorption, and NV$^-$ emission in ensembles of defects~\cite{davies_vacancy-related_1992}. However, ensemble measurements such as photoluminescence, visible and infrared absorption spectroscopy, and electron-spin resonance techniques, do not provide sufficient resolution and lack the sensitivity to probe competing mechanisms driving point defect kinetics (appearances vs disappearances; individual orientation changes). Nor do they typically have the sensitivity to probe ultra-pure diamond samples with very low defect density ($<$ parts-per-trillion) such as those now utilized for quantum information applications. Here we extend this ensemble technique using confocal microscopy, a tool capable of detecting single defects in bulk crystals. Two surprising results relevant to engineering NV centers for applications are obtained: (1) the observation of NV$^-$ density enhancement in the absence of sample irradiation at 950-980\;$^\circ$C and (2) NV$^-$ orientation changes in the absence of full NV dissociation at 1050\;$^\circ$C. \sri{Excitingly, the latter observation points to a route toward engineering the orientation of an NV$^-$ center utilizing strain.} Furthermore, NV quenching, which would be masked in the presence of NV formation in ensemble measurements, suggests NVH$_x$ complexes are a likely candidate for the \sri{spatially homogeneous} disappearances of NV centers at temperatures below 1050\;$^\circ$C. \sri{Thus we find that the longitudinal tracking of single NV$^-$ centers through annealing is a promising tool for studying vacancy diffusion, hydrogen diffusion, the dissociation of hydrogen traps (e.g. NVH$_x$), and partial defect dissociation and re-orientation in ultra-pure diamond.}

\section{Acknowledgements}
This material is based upon work supported by the National Science Foundation under Grants ECCS-180756 (primary for experiment) and DMR-1719797 (primary for theory). We thank undergraduates Brianna Birkel, Kelsey Bates, Phoenix Youngman and Thalya Paleologu for assistance with the optical measurements, Mike Gould for assistance with annealing, and Matthew Markham for helpful discussions. Undergraduate student support for this project was provided under REU supplements to National Science Foundation grants PHY-1150647, ECCS-1506473, EFMA-1640986 and PHY-1559631. 

\bibliography{references, Annealing_kmf}

\end{document}

\title{Supplementary Material: A window into NV center kinetics via repeated annealing and spatial tracking of thousands of individual NV centers }
\author{Srivatsa Chakravarthi}
\email{srivatsa@uw.edu}
\affiliation{Department of Electrical and Computer Engineering, University of Washington, Seattle, Washington 98195, USA}
\author{Chris Moore}
\affiliation{Department of Physics, University of Washington, Seattle, Washington 98195, USA}
\author{April Opsvig}
\affiliation{Department of Electrical and Computer Engineering, University of Washington, Seattle, Washington 98195, USA}
\author{Christian Pederson}
\author{Emma Hunt}
\author{Andrew Ivanov}
\author{Ian Christen}
\affiliation{Department of Physics, University of Washington, Seattle, Washington 98195, USA}
\author{Scott Dunham}
\affiliation{Department of Electrical and Computer Engineering, University of Washington, Seattle, Washington 98195, USA}
\author{Kai-Mei C Fu}
\affiliation{Department of Electrical and Computer Engineering, University of Washington, Seattle, Washington 98195, USA}
\affiliation{Department of Physics, University of Washington, Seattle, Washington 98195, USA}

\date{\today}% It is always \today, today,
\maketitle            

\section{Confocal microscopy apparatus}
\begin{figure}
\centering
\includegraphics[width=3in]{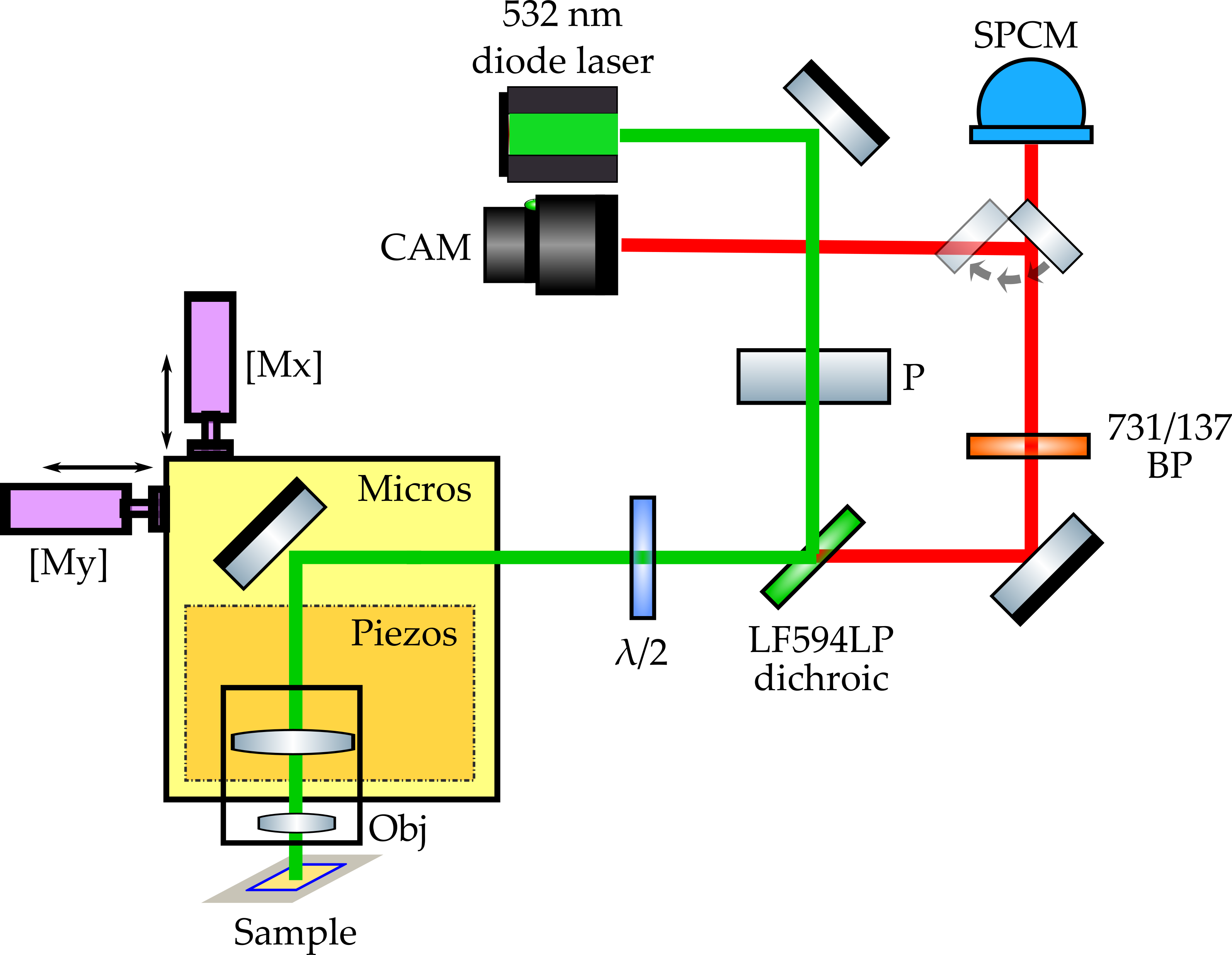}
\caption{Schematic of the microscope setup}
\label{fig:microscope}
\end{figure}

\begin{figure}[]
\centering
\includegraphics[width=4in]{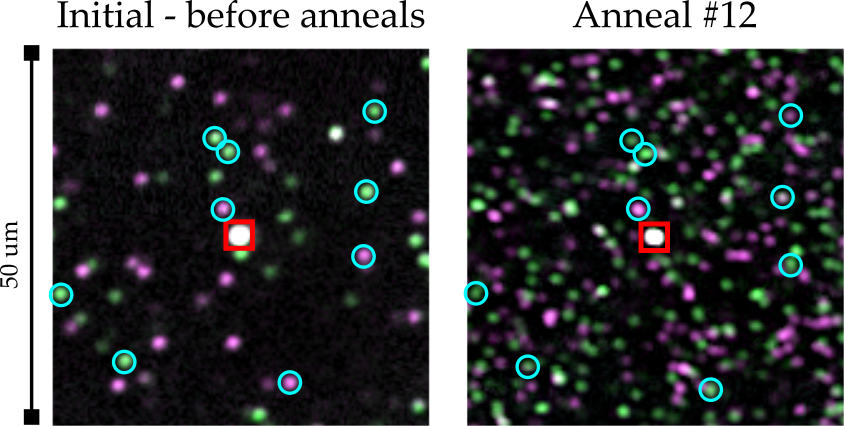}
\caption{Crop (50~\textmu m$~\times~$50~\textmu m) of large-area scan data from Sample B showing the increase in NV$^-$ density from initial (no-anneals) state to post anneal \#12 (maximum NV$^-$ density). The blue circles mark NVs retained through the anneals. The red square marks a persistent defect used for image registration.}
\label{fig:sampleB}
\end{figure}

Optical measurements were performed using a home-built confocal microscope (see Fig.~\ref{fig:microscope}) with a 0.75 NA, 40x objective (Nikon Fluor 40X). A piezo stage (MadCityLabs nano-positioner) was used for imaging 50x50~\textmu m$^2$ area scans (see Fig.~\ref{fig:sampleB}), while a micrometer stage comprised of two 25~mm actuators (Newport TRA25CC) was used to move to sequential scanning locations in steps of 40~\textmu m. A 637~nm laser reflection signal was used between subsequent piezo scans to track the sample surface to maintain the experimental depth to within 2~\textmu m. The nitrogen-vacancies were optically excited with a 532~nm diode-pumped solid-state laser (RGBLase LLC FB-532-100-FS-FS-010-1N) at powers between 4-5~mW \sri{with a spot diameter of $\sim$1.6~\textmu m. The spot size was chosen to increase the number of NV$^-$ centers within the depth of focus. The microscope point spread function is an ellipsoid with axes 1.8~\textmu m, 1.5~\textmu m, and (10$\times$2.4)~\textmu m}. A linear polarizer (ThorLabs LPVIS100) was used in the excitation path and a half-wave plate was used to preferentially excite a given NV$^-$ orientation. NV$^-$ phonon-sideband emission was filtered into the collection path by a combination of dichroic beam splitter (Semrock LF594LP) and a 660-800~nm bandpass filter (Semrock FF01-731/137-25). The NV$^-$ photons were measured with an avalanche photodiode (Excelitas SPCM-AQ4C). Photoluminesence from Sample C was attenuated with an optical density filter (Newport-530-OD3 or Newport-530-OD2) to avoid detector saturation. \sri{NV$^-$ density for samples A and B were calculated by counting total NV$^-$ in the large-area scan data and normalizing by the scan area.} NV$^-$ density calibration for the sample C was performed by measuring the point-spread function and count-rate of a single NV$^-$. Calibration measurements were performed after each anneal to account for microscope changes. 

\section{Data Analysis Techniques}

\begin{figure}
\centering
\includegraphics[width=5in]{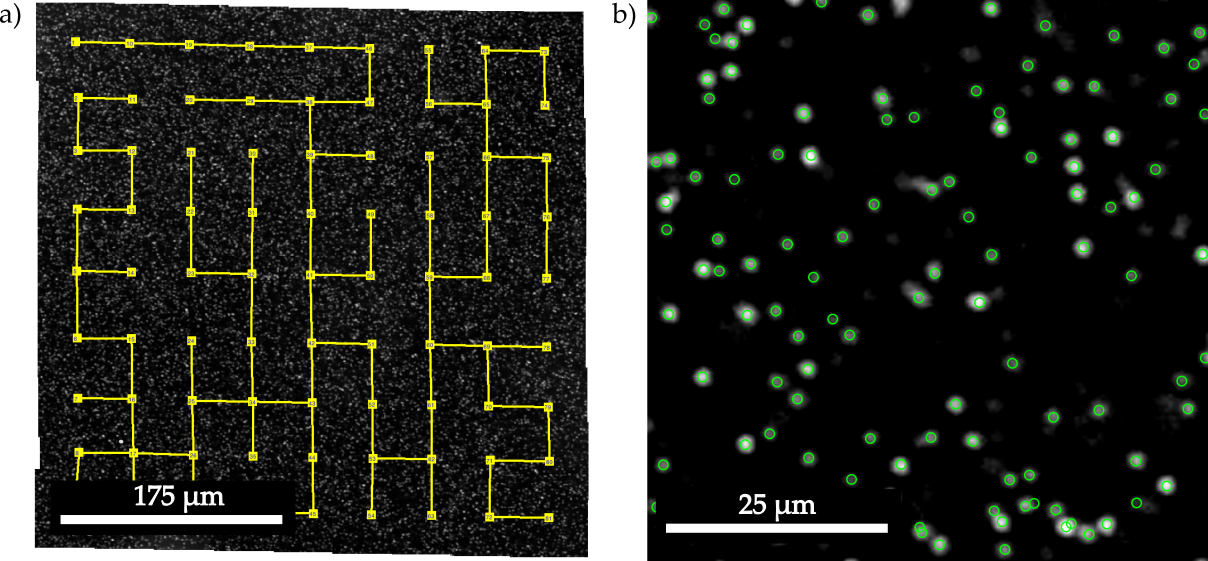}
\caption{(a) $\SI{350}{\micro\metre}\times\SI{350}{\micro\metre}$ stitched large-area scan after anneal \#12 showing the MST graph. (b) A $\SI{50}{\micro\metre}\times\SI{50}{\micro\metre}$ scan showing NV$^-$ position detection via CHT algorithm.}
\label{fig:data_processing}
\end{figure}

{\it Large-area scan stitching}:
large-area scans, composed of $9\times9$ overlapping
$\SI{50}{\micro\metre}\times\SI{50}{\micro\metre}$ scans for each of the two distinguishable orientations, were stitched into composite images with a custom tool. First, the tool performed normalization across all individual scans to ensure uniform appearance of NV intensities in the output images; then it calculated pairwise alignments (via digital image correlation with subsequent brute-force optimization) for all adjacent area scans along with corresponding misalignment scores. The final composite image was produced by selecting pairwise alignments forming a minimum spanning tree (MST, see Fig.~\ref{fig:data_processing}(a)) with respect to the misalignment scores. This fully-automated process generally produced quality results; however, in order to compensate for possible imperfections, the tool also allowed manual adjustments of pairwise alignments in the MST. The tool is publicly available on github (https://github.com/optospinlab/lateral-scan-analysis).

\vspace{1mm}
{\it Registration of large-area scans}:
Successive pairs of large-area scans were aligned using affine homography transformation (\sri{performed by \emph{ImageJ}~\cite{ref:rueden2017iif,ref:schindelin2012fao}}). Essentially, this transformation amounted to a rotation and perspective transformation such that NV$^-$ centers present in both scans overlapped, which compensated for discrepancies in the orientation and tilt of the sample under the microscope. The overlapping regions of aligned pairs of successive scans were then compared by means of image subtraction, which allowed manual and automated detection of the changes in the spatial maps of the NV$^-$ centers. 

\vspace{1mm}
{\it Tracking NV$^-$ changes}:
The registered large-area scan images from consecutive anneals (say anneal A \& B) are processed through an image subtraction algorithm after normalization. First the data sets A \& B are converted to 3D matrices, with the array indices indicating \{position x , position y, orientation (1 or 2)\} and the array values representing the intensity (measured photon counts). Each individual scan constituting matrices A \& B were normalized prior to stitching, however excitation laser power and collection efficiency can vary slightly between anneals. Thus NV$^-$ intensity histograms of the two data sets (A \& B) are used for normalization (factor N) between anneals. This allows us to clearly identify and subtract the background during comparison. The following matrix algebra was performed to extract appearance, disappearance and orientation-change information:

\begin{align}
\mathrm{C} &= \mathrm{A(x,y,1)}/\mathrm{N} - \mathrm{B(x,y,1)}\\
\mathrm{D} &= \mathrm{A(x,y,2)}/\mathrm{N} - \mathrm{B(x,y,2)}\\[10pt]
\mathrm{E} &= \mathrm{C} + \mathrm{D}\\
\mathrm{F} &= \mathrm{C} - \mathrm{D}
\end{align}
\begin{align}
\mathrm{Appearances(x,y)} &=
\begin{cases}
\mathrm{-E} & \text{if } \mathrm{E} < 0\\
0 & \text{if } \mathrm{E} \geq 0\\
\end{cases}\\
\mathrm{Disappearances(x,y)} &=
\begin{cases}
\mathrm{E} & \text{if } \mathrm{E} \geq 0\\
0 & \text{if } \mathrm{E} < 0\\
\end{cases}\\
\mathrm{Orientation~changes(x,y)} &= \mathrm{abs(~abs(F)-abs(E)}\times \mathrm{F)}
\end{align}\\
\\

For visualization of differences, the three matrices; Appearances, Disappearances and Orientation changes, represented by blue, grey and yellow colors respectively, are combined into a single RGB image to generate the difference image, shown in Fig.~2. The next step after the identification of differences is to map the spatial location (x,y) of every NV with the corresponding changes, if any. This allows us to track the evolution of every individual NV$^-$ defect through all the anneals. This is performed by MATLAB code implementing the Circular Hough Transform (CHT) to detect NV$^-$ centers and record their centroid (see Fig.~\ref{fig:data_processing}(b)). This code also provides the total number of NV$^-$ centers and differences reported in Fig. 3 of the main text. A similar code is used to process the depth scan data. The processing code and raw data are available here upon request.  \\   

\vspace{1mm}
\sri{{\it Conversion from NV$^-$ number/area to density}: While each large-scan area was nominally (360~\textmu m)$^2$, due to an image offset, image rotation, or failed stitching, only a subset of this area could be reliably matched before and after each anneal. This cropped area was utilized, in addition to the 25~\textmu m confocal depth of focus, to obtain the NV$^-$ density used in Table I and Fig. 3 in the main text. Table~\ref{tab:countingdata} provides the NV$^-$ counting data and crop areas for sample B.}\\

\vspace{1mm}
\sri{{\it Uncertainty estimate in total NV$^-$ density in Fig.~3 of the main text}. For each anneal, this estimate consists of two components, E$_1$ and E$_2$, added in quadrature.  E$_1$ represents error due to the automated image analysis. This error is estimated by comparing the change in the total NV's counted before and after each anneal to the difference in appearances and disappearances detected. E$_1$ ranges from 2-10\%. The second uncertainty E$_2$ arises from the density variation across the scan area (Figs.~\ref{fig:nvdist}(a),(b)) combined with the different comparison areas used. Assuming a random density variation with the magnitude given by Fig.~S4 and the crop-areas given in Table S1, this density variation results in an estimated uncertainty of 4\%.}     \\

\begin{table*}[t]
\noindent
  \begin{tabular} {|| m{3cm} | m{0.9cm}m{0.9cm}m{0.9cm}m{0.9cm}m{0.9cm}m{0.9cm}m{0.9cm}m{0.9cm}m{0.9cm}m{0.9cm}m{0.9cm}||}
  \hline
    Anneal \# & 1-3 &  4,5 & 6,7 & 8 & 9, 10 & 11 & 12 & 13 & 14 & 15 & 16 \\
    \hline\hline
   \rowcolor[rgb]{0.8, 0.8, 0.8}[2pt][2pt] 
   Crop Area (\textmu m)$^2$  & (282)$^2$ & (287)$^2$& (285)$^2$& (299)$^2$& (307)$^2$& (298)$^2$& (291)$^2$& (339)$^2$& (331)$^2$& (303)$^2$& (311)$^2$ \\
   Total (post-anneal) & 2126 & 2822 & 3213 & 4635 & 5579 & 8772 & 9174 & 11511 & 10080 & 8225 & 8456 \\
    \rowcolor[rgb]{0.8, 0.8, 0.8}[2pt][2pt]
    Appearances & 732 & 1101 & 672 & 1014 & 664 & 3641 & 1266 & 543 & 122 & 56 & 70 \\
   Disappearances & 145 &  602 & 47 & 41 & 19 & 80 & 157 & 618 & 1478 & 90 & 181 \\
    \rowcolor[rgb]{0.8, 0.8, 0.8}[2pt][2pt]
    Reorientations & 6 & 17 & 1 & 6 & 4 & 64 & 129 & 272 & 1163 & 1642 & 3093 \\
    \hline
  \end{tabular}
      \caption{Sample B NV$^-$ data used in Fig.~3 and Table I in the main text.}
    \label{tab:countingdata}
\end{table*}

\section{Sample orientation}

\sri{During imaging, the ``top'' of the sample refers to the surface of the sample closest to the microscope objective. The ``bottom'' of the sample rests on a silicon chip. During annealing, the sample was placed in a sapphire boat with one side of the sample in contact with the boat surface and the other exposed to the vacuum. In all anneals except Sample B, Anneal 15, the bottom of the sample coincided with the surface in contact with the boat. During Anneal 15, the sample was flipped.}  

\section{Variation of substitutional nitrogen across sample B}
\begin{figure}
\centering
\includegraphics[width=5.8in]{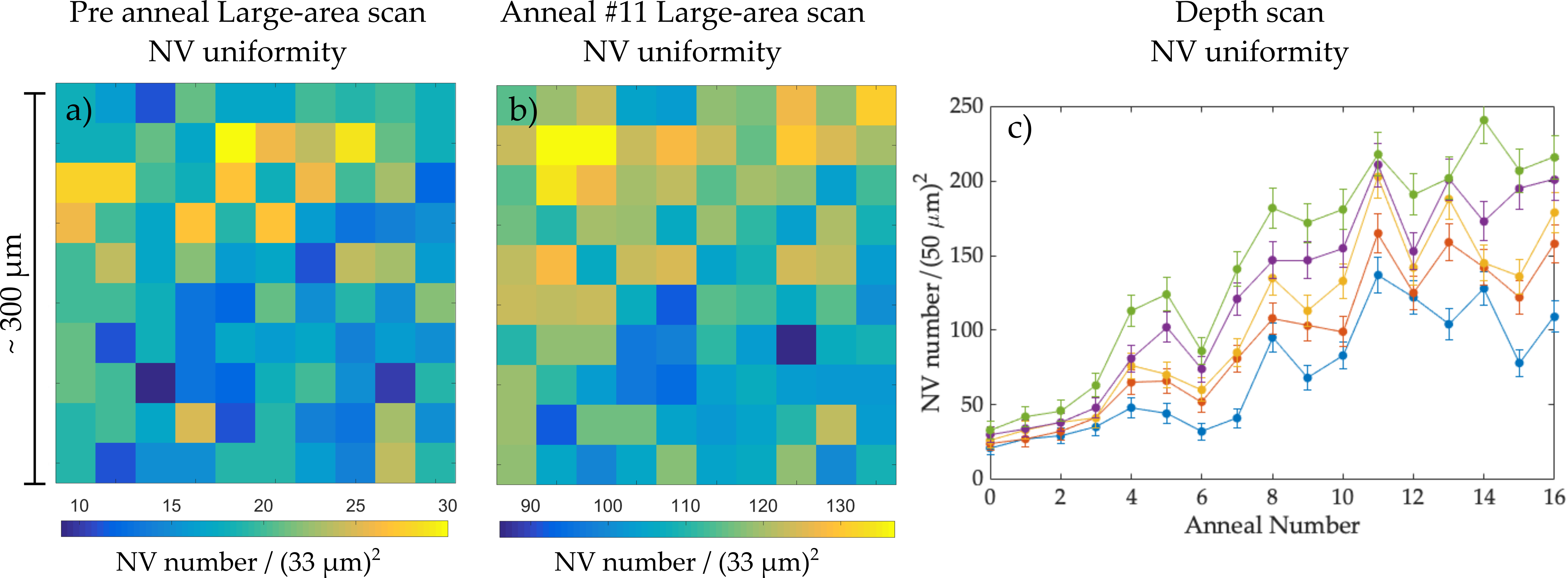}
\caption{(a) Variation in NV$^-$ density across large-area scan region pre-anneal. (b) Variation in NV$^-$ density across large-area scan region after anneal\#11. (c) Variation in NV$^-$ density across 5 sampled regions at a depth of 240~\textmu m, as a function of anneal number. Each location is denoted by a different color. Error bars indicate one-sigma uncertainty for a Poisson distribution.}
\label{fig:nvdist}
\end{figure}

Fig.~\ref{fig:nvdist}(a,b) illustrates the inhomogeneity of the NV$^-$ distribution across the sample. When taking depth scan data, 5 locations approximately 500~\textmu m apart from each other were utilized. Fig.~\ref{fig:nvdist}(c) plots the NV$^-$ density in each of the 5 locations (corresponding to the different colors) at a depth of 240~\textmu m. Note that between anneals, the locations corresponding to the same color are not exactly the same but can differ by up to $\approx$100~\textmu m. We can see that the NV$^-$ density can vary by up to a factor 3 at a given depth.

\section{Data from samples A, D, E}

\begin{figure}[h]
\centering
\includegraphics[width=5in]{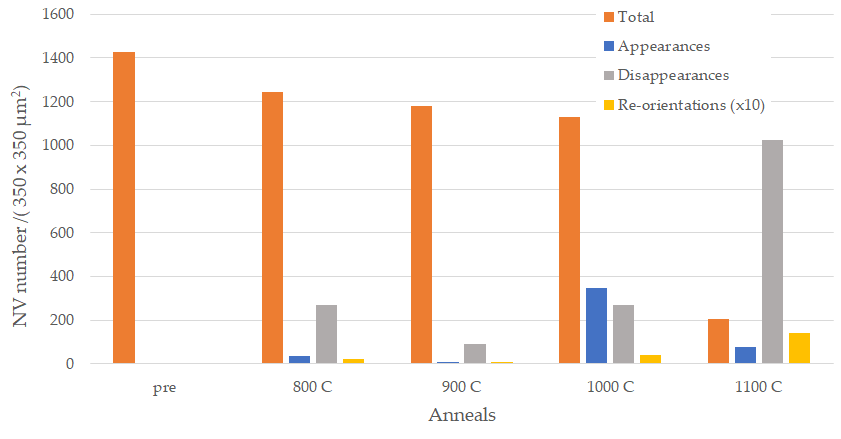}
\caption{Large-area scan data from sample A at a depth of 96~\textmu m, as a function of anneal temperature. The ramp and hold times were each 2 hours for every anneal. The appearances and disappearances were manually counted in this dataset.}
\label{fig:sampleA}
\end{figure}

\begin{figure}[h]
\centering
\includegraphics[width=5.8in]{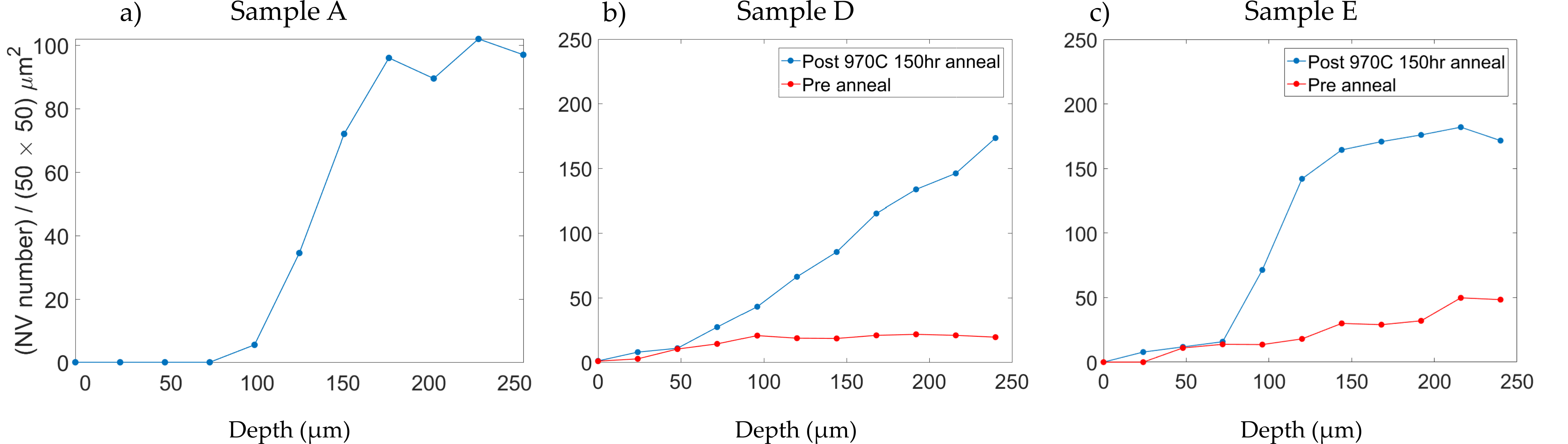}
\caption{(a) Number of NVs in Sample A as a function of depth after 1100$^\circ$C anneal. (b) Sample D pre and post anneal. (c) Sample E pre and post aneneal.}
\label{fig:sampleAdepth}
\end{figure}

\sri{Here we include additional data from electronic-grade samples A, D, and E. Sample A is the initial sample in which single NV$^-$ centers are tracked before and after annealing.  In Fig.~\ref{fig:sampleA}, disappearances dominate in both the 800$^\circ$C and 900$^\circ$C anneals. This is consistent with what is observed in Sample B for these relatively short, 2 hour anneals. Appearances are observed at 1000$^\circ$C. At 1100$^\circ$C nearly all NV$^-$ centers at the 96~\textmu m test depth are depleted. To test whether this is due to an inhomogeneous surface-driven process, NV$^-$ center density was measured as a function of depth. As shown in Fig.~\ref{fig:sampleAdepth}(a), the disappearances appear to be related to the surface. Depth scan data before and after the single 980\;$^\circ$C 150~h anneal for samples D and E} also exhibit this surface depletion effect as shown in  Figs.~\ref{fig:sampleAdepth}(b,c).

After these measurements, the top surface of Sample A was etched 4~\textmu m via argon/oxygen plasma RIE and then implanted with $^{15}$N (85keV and 1.4e10/cm$^2$ dose) for another experiment. Several electronic grade samples which had not undergone anneals were implanted during the same run. In contrast to the other implanted samples, in Sample A, we did not observe any formation of NV centers (neither NV$^0$ nor NV$^-$ charge state) after vacuum annealing (temperature profile: 2~hr at 400\;$^\circ$C; 8~h at 800\;$^\circ$C; 2~h at 1100\;$^\circ$C with 2~h ramps). This could indicate formation of optically inactive NVH$_x$ or V$_x$H$_y$ complexes due to hydrogen introduced during the prior 1100\;$^\circ$C 2~h anneal.

\begin{figure}[h]
\centering
\includegraphics[width=5.8in]{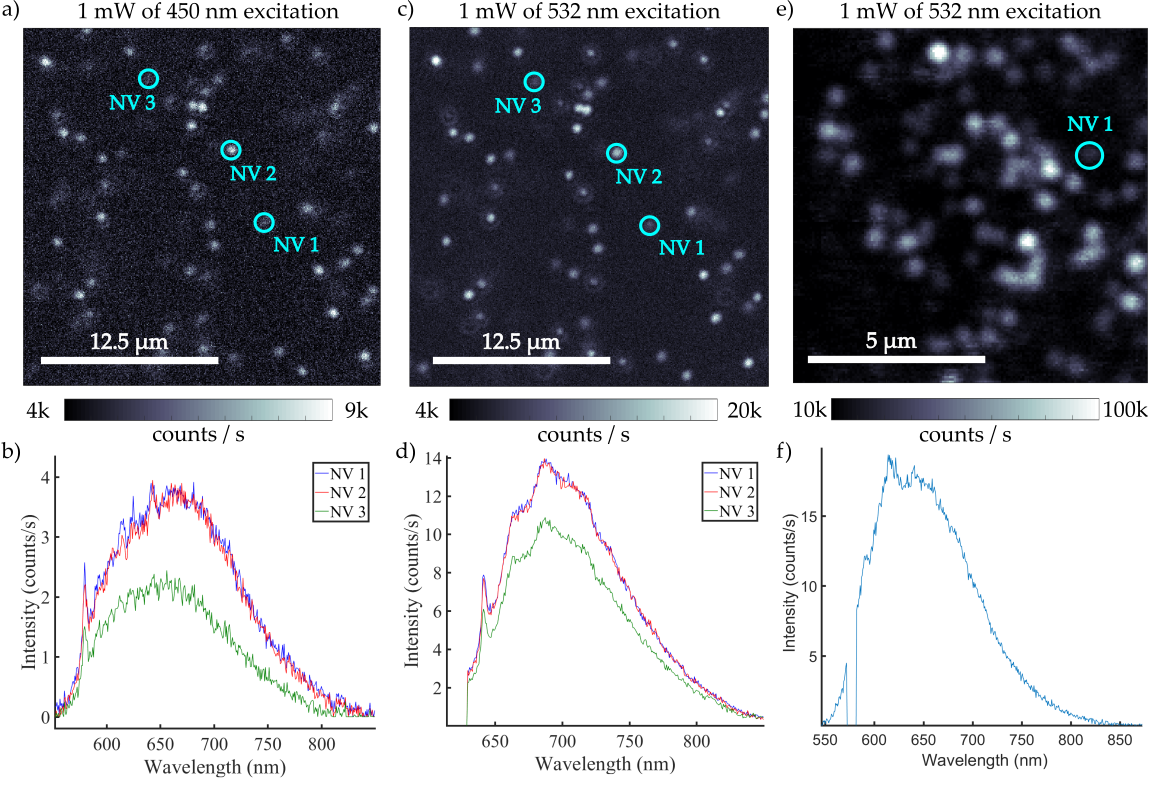}
\caption{\sri{(a,b) Confocal scan and spectra of 3 NVs in sample B taken with 450~nm excitation. (c,d) Confocal scan and spectra of same 3 NVs in sample B taken with 532~nm excitation. (e,f) Confocal scan and spectra of NV$^0$ from a reference sample taken with 532~nm excitation.}}

\label{fig:nv_charge_state}
\end{figure}

\section{NV charge state identification}

\sri{We consider charge state conversion from the bright NV$^-$ charge state into the dark NV$^0$ as a possible explanation for NV disappearances in our scans by investigating the depletion region of sample B. ~\ref{fig:nv_charge_state}(a,c) show confocal scans taken of the same region with 1~mW of 450~nm and 532~nm excitation respectively. Collection in ~\ref{fig:nv_charge_state}(a) includes wavelengths greater than 539~nm while ~\ref{fig:nv_charge_state}(c) includes only collection above 600~nm. 532~nm excitation is commonly used to detect NV centers, however the relatively bright first (575~nm) and second (611~nm) order diamond Raman lines occur in the NV$^0$ PL band, which could obscure the weak PL from a single NV$^0$. We utilize 450~nm excitation to solve this issue since the first order (479~nm) and second order (511~nm) Raman lines are no longer collected. If charge state conversion were to be responsible for the disappearances, we would expect to see defects in the 450~nm confocal scan that did not appear in the 532~nm confocal scan, but this is not observed.

~\ref{fig:nv_charge_state}(d) shows characteristic, background subtracted spectra of 3 NVs in the scan region under 532~nm excitation, resulting from the NVs existing in the ratio of NV$^0$  to NV$^-$ as 1:3. NV-3 is dimmer than the other NVs due to the orthogonal polarization.  In ~\ref{fig:nv_charge_state}(b) we only observe NV$^0$ PL from the same 3 defects under 450~nm excitation, indicating that NV$^-$ ionizes without contributing to the PL. We are thus able to observe single NV$^0$. On these grounds we exclude charge state conversion as a cause for disappearances. For reference, spectra from a sample containing predominantly NV$^0$ is shown in ~\ref{fig:nv_charge_state}(e,f)}

\section{Estimation of NV reorientation barrier}
The reorientation rate was calculated for anneals 11-16 using the data in Table~\ref{tab:countingdata} and the 150 hour anneal time. We neglect the effect of appearances and disappearances and the possibility of multiple reorientation cycles in a single anneal. The fractional change between two distinct orientations, e.g. orientations 1 changing to 2, is $F_{12} = \Delta N_{12}/N_1$ in which $\Delta N_{12}$ is the number of changes from orientation 1 to 2 and $N_1$ is the population of orientation 1. We can approximate\sri{ $\displaystyle F_{12} = \Delta N_{12}/N_1 = \Delta N_\textrm{total, exp}/(N_\textrm{total})$}, in which $N_\textrm{total, exp}$ is the total number of reorientations experimentally detected and $N_\textrm{total}$ is the total number of NV$^-$ centers. This approximation assumes all orientations are equally populated and all orientation changes are equally probable. In calculating the fractional change, we utilized the population of the NV$^-$ centers before the anneal. To calculate the rate, we divide the fractional change by the annealing time.

\begin{figure}[h]
\centering
\includegraphics[width=3in]{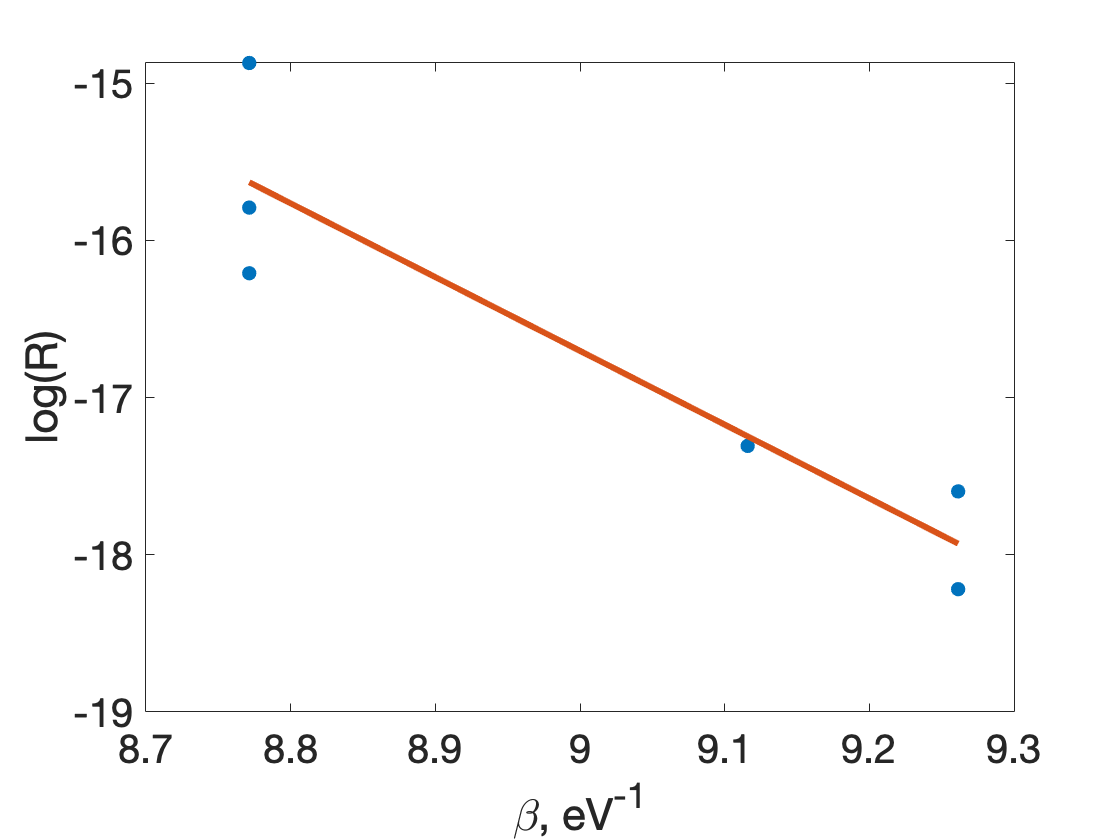}
\caption{Log of the reorientation rate as a function of $\beta = (k_B T)^{-1}$}
\label{fig:orientationchanges}
\end{figure}

The rate for a single orientation change, {\it e.g.} $R_{12}$ from orientation 1 (green) to 2 (purple), will be given by $R_{12} = F_{12}/t = \nu~\mathrm{exp}(-E_{b,12}/kT)$ where t is the anneal duration and T is the anneal temperature in K.  In Fig.~\ref{fig:orientationchanges}, $\log(R_{12})$ is plotted as a function of $\beta = (k_BT)^{-1}$ and is fit to a line. The absolute value of the slope determines the re-orientation barrier $E_b$ and the x-intercept determines $\log(\nu)$, which $\nu$ is an effective attempt frequency. A least square fit gives $E_b = 4.7 \pm 0.9 eV$ and $\nu=\exp(26\pm9)~\textrm{s}^{-1}$. Both the calculated value of $E_b$ (4.85 eV~\cite{ref:pinto2012odn} and a typically used attempt frequency of 30 THz lie in the uncertainty range.

%\bibliography{Annealing-kmf}

%Control: production of eprint (-1) disabled
%